\documentclass[aps,pra,twocolumn,showpacs,showkeys,amsmath,superscriptaddress,floatfix]{revtex4-1}
\usepackage{graphicx}
\usepackage{epsfig}
\usepackage{dcolumn}
\usepackage{bm}
\usepackage{amsmath}
\usepackage{amsbsy}
\usepackage{amssymb}
\usepackage[usenames]{color}
\usepackage{multirow}
\usepackage{soul}
\usepackage[normalem]{ulem}

\newcommand{\bra}[1]{\ensuremath{\langle{#1}|\,}}
\newcommand{\ket}[1]{\ensuremath{\,|{#1}\rangle}}

\renewcommand{\vr}{\textbf{r}}
\newcommand{\vE}{\textbf{E}}
\newcommand{\vB}{\textbf{B}}
\newcommand{\vA}{\textbf{A}}

\begin{document}


\title{Formulation of the twisted-light--matter
interaction at the phase singularity: \\
beams with strong magnetic fields}

\author{G. F. Quinteiro}

\email{gquinteiro@df.uba.ar}

\affiliation{ Departamento de F\'isica and IFIBA, FCEN, Universidad de Buenos
Aires, Ciudad Universitaria, Pabell\'on I, 1428 Ciudad de Buenos Aires,
Argentina}

\affiliation{Universit\"at M\"unster, Wilhelm-Klemm-Str. 10, 48149 M\"unster,
Germany}

\author{D. E. Reiter}

\affiliation{Universit\"at M\"unster, Wilhelm-Klemm-Str. 10, 48149 M\"unster,
Germany}

\author{T. Kuhn}

\affiliation{Universit\"at M\"unster, Wilhelm-Klemm-Str. 10, 48149 M\"unster,
Germany}

\date{\today}

\begin{abstract}
The formulation of the interaction of matter with singular light
fields needs special care. In a recent article [Phys.~Rev.~A {\bf
91}, 033808 (2015)] we have shown that the Hamiltonian describing
the interaction of a twisted light beam having parallel orbital and
spin angular momenta with a small object located close to the phase
singularity can be expressed only in terms of the electric field of
the beam. Here, we complement our studies by providing an
interaction Hamiltonian for beams having antiparallel orbital and
spin angular momenta. Such beams may exhibit unusually strong
magnetic effects. We further extend our formulation to radially and
azimuthally polarized beams. The advantages of our formulation are
that for all beams the Hamiltonian is written solely in terms of the
electric and magnetic fields of the beam and as such it is
manifestly gauge-invariant. Furthermore it is intuitive by
resembling the well-known expressions in the dipole-electric and
dipole-magnetic moment approximations.
\end{abstract}

\maketitle


\section{Introduction}

Typically, when studying the interaction of light with nanometer-sized
structures the characteristic length scale of the light field is much larger
than the size of the structure. In this case it is usually sufficient to
consider plane wave-like or spatially homogeneous beams. This does not hold
anymore if the structure is placed at or close to a singular point of a light
beam.

A prominent example for such a singular light beam is twisted light (TL),
also called optical vortex light or light carrying orbital angular momentum,
which has a phase singularity at the beam axis.
A variety of new effects have been predicted and observed in the
study of TL beams, spanning pure optics \cite{andrews2011str,
ballantine2016there, yamane2012ultrashort} and the interaction with
atoms \cite{koksal2012cha, surzhykov2015interaction}, molecules
\cite{watzel2016optical}, ions \cite{schmiegelow2012light,
peshkov2015ionization}, Bose-Einstein condensation
\cite{bhowmik2016interaction}, and solid-state systems
\cite{quinteiro2009the,shigematsu2013orb,
clayburn2013sea,noyan2015time,shintani2016spin, kocc2015quantum}.
All these effects promise interesting new applications to material
processing \cite{omatsu2010metal}, communications
\cite{spinello2016radio}, lasers \cite{zhang2016self,
abulikemu2016octave,miao2016orbital}, spintronics
\cite{quinteiro2014light} and particle manipulation
\cite{woerdemann2013advanced}.

Another class of spatially strongly inhomogeneous light beams are
radially and azimuthally polarized beams, which can be realized as
linear combinations of TL beams with opposite angular momentum and
circular polarization. These beams have received much attention for
their high potential in applications. Thanks to their strong
longitudinal-field component with high intensity and degree of
focusing, they prove useful in fields like micro-Raman spectroscopy
\cite{saito2008zpolarization}, material processing
\cite{wang2008creation,meier2007material}, and as optical tweezers
for metallic particles \cite{zhan2004trapping}. It was also
suggested that a strong longitudinal component can help to excite
intersubband transitions in quantum wells \cite{sbierski2013twisted}
and light-hole states in quantum dots \cite{quinteiro2014light}.
These states are technologically challenging to address, since
conventional fields can only excite them if the beam propagates
perpendicular to the growth direction of the sample, which typically
requires cleaving the structure. From a theoretical perspective it
has been also demonstrated that these fields can be classically
entangled in a way similar to what we find in quantum mechanical
systems \cite{gabriel2011entangling}.

It is becoming increasingly clear that the interaction of highly
inhomogeneous light fields, and in particular of singular fields like TL
\cite{dennis2009singular}, with atoms or solids is non-trivial and needs
special care in the theoretical description. Here, the widely used
dipole-moment approximation cannot be applied anymore. Of course, one can
always work with the minimal coupling Hamiltonian; however, its use entails
some disadvantages, for example it lacks direct connection to the
electro-magnetic fields, the real quantities accessible in experiments. In
Ref.~\cite{quinteiro2015formulation}, we have shown recently that the
formulation of the light-matter interaction had to be revisited and
demonstrated that previous formulations meant for smooth fields are not the
most suitable ones to treat TL, especially when the interaction with small
structures close to the phase singularity is considered. Using elementary
gauge transformations, we further developed a new gauge --the TL gauge--
which allowed us to cast the Hamiltonian in a form containing the electric
field only.

\begin{figure}[h]
  \centerline{\includegraphics[scale=.3]{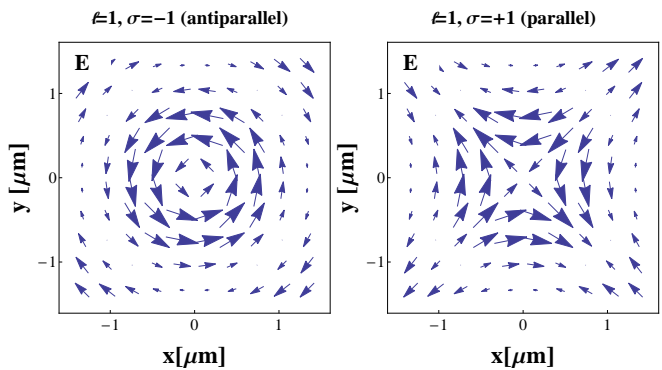}}
  \caption{Electric field profile in the $xy$-plane for parallel and antiparallel
  TL propagating along the $z$-direction. We chose the topological charge to $\ell=1$
  and the handedness of polarisation to $\sigma=1$ (left) and $\sigma=-1$ (right).}
  \label{fig:Profile}
\end{figure}
Though having an appealing form, the TL gauge developed in
Ref.~\cite{quinteiro2015formulation} is only applicable to a certain subclass
of TL beams, which can be explained as follows: TL beams can be discriminated
into two topologically different classes depending on the combination of
circular polarization (or spin angular momentum) and topological charge (or
orbital angular momentum). If circular polarization and topological charge
have the same sign, we call this the parallel class, while for opposite sign
the beams are called antiparallel. One example of the electric field profile
for the two classes is shown in Fig.~\ref{fig:Profile}. One can immediately
see the difference in the spatial profiles of the beams, which even by
evolving in time will not transform into each other. Coming back to the
TL-matter interaction, we have shown that the TL gauge can only be applied to
the parallel class \cite{quinteiro2015formulation} since it does not account
for a magnetic coupling, which turns out to be crucial in the antiparallel
class.

In this paper we extend the description of the TL-matter interaction
to the family of antiparallel TL beams by including both electric
and magnetic interaction terms. We will show that for antiparallel
beams with orbital angular momentum larger than one the magnetic
interaction becomes unusually strong. The formulation can also be
directly applied to the interaction of radially or azimuthally
polarized beams with small structures close to the beam center. The
light-matter Hamiltonian derived here shares the benefits of our
previous TL gauge, namely, it is intuitive and easy to use. Thereby,
this article completes the gauge invariance formulation of the
TL-matter interaction close to the phase singularity.

We organize the article as follows. Section \ref{Sec:OV} describes the modes
of twisted light, providing the expressions for electric and magnetic fields
close to the phase singularity. Like in our previous paper we will restrict
the explicit formulas to Bessel modes. Since the derivation only relies on
the behavior close to the phase singularity, however, the general features
are also valid for other types of beams like, e.g., Laguerre-Gauss (LG)
beams. The derivation and formulation of the TL--matter Hamiltonian in terms
of electric and magnetic fields is given in Sect.~\ref{Sec:Formulation}
followed by a discussion of the resulting Hamiltonain in
Sect.~\ref{Sec:diss}. Section \ref{Sec:Ext} treats the case of radially  and
azimuthally polarized fields. The conclusions are presented in
Sect.~\ref{Sec:Conc}.

\section{Bessel singular fields}
\label{Sec:OV}

The most significant feature of TL is its topological charge $\ell$ that adds
orbital angular momentum via the phase $\exp(i \ell \varphi)$, where
$\varphi$ is the angle in the cylindrical coordinates $\{r, \varphi, z\}$ for
a beam centered around $r=0$ and propagating in the $z$-direction. This
implies a phase singularity at $r=0$ whenever $\ell \neq 0$. Another
important parameter is the handedness of the circular polarization denoted by
$\sigma=\pm1$. The combination of the signs of $\ell$ and $\sigma$ leads to
the distinction into the parallel [sign($\sigma$)=sign($\ell$)] and the
antiparallel [sign($\sigma$) $\ne$ sign($\ell$)] classes. In the radial modes
one distinguishes between LG and Bessel modes. The main difference between
these types is their radial localization, i.e., their behavior for large
values of $r$. Close to the beam axis they behave similarly. In this paper we
will restrict ourselves to the case of Bessel beams because they are exact
solutions of the full Helmholtz equation \cite{volke2002orb} and therefore
can be applied also beyond the limits of the paraxial approximation.
Furthermore they are non-diffracting beams, such that the radial profiles are
independent of the propagation coordinate $z$.

Bessel beams can be derived from the vector potential in the Coulomb gauge,
as explained in App.~\ref{App:Bessel}. We are interested in the description
of the light-matter interaction close to the phase singularity. Thus, we
approximate the full fields given in App.~\ref{App:Bessel} in the region $q_r
r \ll 1$, where $1/q_r$ is a measure of the beam radius. This is basically
done by expanding the Bessel functions $J_{\ell}(q_r r) \propto (q_r
r)^{|\ell|}$ \cite{quinteiro2015formulation}. To simplify the notation here
we will assume $\ell > 0$. The extension to negative values is
straightforward. Note that the formulas for the full fields given in
App.~\ref{App:Bessel} hold for arbitrary values of $\ell$.

Separating the propagating phase from the mode functions according to
$\vE(\vr,t)=\frac{1}{2} \tilde{\vE}(\vr)e^{i(q_z z-\omega t)} +
\textrm{c.c.}$ and $\vB(\vr,t)=\frac{1}{2}  \tilde{\vB}(\vr)e^{i(q_z z-\omega
t)} + \textrm{c.c.}$, where $\textrm{c.c.}$ denotes the complex conjugate,
the electric field for circular polarization $\sigma=\pm 1$ reads
\begin{subequations}
\label{Eq:E_Bessel}
\begin{eqnarray}
\label{Eq:EBx}
    \tilde{E}_x(\mathbf r)
&=&
    i  \frac{E_0}{2^\ell \ell!}  (q_r r)^\ell e^{i \ell \varphi}
\\
\label{Eq:EBy}
    \tilde{E}_y(\mathbf r)
&=&
    -\sigma   \frac{E_0}{2^\ell \ell!}  (q_r r)^\ell
    e^{i \ell \varphi}
\\
\label{Eq:EBz}
    \tilde{E}_z(\mathbf r)
&=&
    \sigma
    \frac{E_0}{2^{\ell+\sigma} (\ell+\sigma)!}
    \frac{q_r}{q_z}
    (q_r r)^{\ell+\sigma}
    e^{i (\ell+\sigma) \varphi}
    \hspace{7 mm}
\end{eqnarray}
\end{subequations}
with the electric field amplitude $E_0$, the frequency $\omega$, and the wave
vector components $q_z$ and $q_r$. The latter quantities are related by
$q_z^2+q_r^2 = (n\omega/c)^2$, $n$ being the index of refraction of the
medium.

From Eqs.~(\ref{Eq:E_Bessel}) we already notice a qualitative difference
between the parallel and the antiparallel class: While in the parallel class
the electric field close to the origin is dominantly in-plane, in the
antiparallel class the $z$-component becomes dominant. The differences are
even more pronounced in the case of the magnetic field which for $\sigma =
+1$ (parallel class) is given by
\begin{subequations}
\label{Eq:B_Bessel_par}
\begin{eqnarray}
\label{Eq:B_x_par}
    \tilde{B}_x(\mathbf r)
&=&
    \frac{B_0}{2^\ell \ell !}
    \left[
        1 + \frac{1}{2}\left(\frac{q_r}{q_z}\right)^2
    \right] \left( q_r r \right)^{\ell}
    e^{i \ell \varphi}  \hspace{7 mm}
\\ \label{Eq:B_y_par}
    \tilde{B}_y(\mathbf r)
&=&
    i \frac{B_0}{2^\ell \ell !}
    \left[
        1 + \frac{1}{2}\left(\frac{q_r}{q_z}\right)^2
    \right] \left( q_r r \right)^{\ell}
    e^{i \ell \varphi}
\\ \label{Eq:B_z_par}
    \tilde{B}_z(\mathbf r)
&=&
    - i \frac{B_0}{2^{\ell+1} (\ell+1)!}
    \frac{q_r}{q_z}
    \left( q_r r \right)^{\ell+1}
    e^{i (\ell+1) \varphi}
\,,
\end{eqnarray}
\end{subequations}
with $B_0 = (q_z / \omega) E_0$, while for $\sigma=-1$ (antiparallel class)
it reads
\begin{subequations}
\label{Eq:B_Bessel_antipar}
\begin{eqnarray}\label{Eq:B_x_antipar}
    \tilde{B}_x(\mathbf r)
&=&
  - \frac {B_ 0} {2^\ell \ell!}
  \left\{
    \left[
        1 + \frac {1} {2}\left (\frac {q_r} {q_z} \right)^2
    \right]
    - \frac {\ell} {2} \left (\frac {q_r} {q_z} \right)^2
    e^{-i 2 \varphi}
  \right.
\nonumber \\
&&
  \left.
    + \frac {1} {2} \left (\frac {q_r} {q_z} \right)^2
    \frac{4\ell (\ell - 1)}{(q_r r)^{2}} e^{-i 2 \varphi}
  \right\}
  (q_r r)^{\ell} e^{i \ell \varphi}
\\ \label{Eq:B_y_antipar}
    \tilde{B}_y(\mathbf r)
&=&
  i \frac {B_ 0} {2^\ell \ell!}
  \left\{
    \left[
        1 + \frac {1} {2}\left (\frac {q_r} {q_z} \right)^2
    \right]
    + \frac {\ell} {2} \left (\frac {q_r} {q_z} \right)^2
    e^{-i 2 \varphi}
  \right.
\nonumber \\
&&
  \left.
    - \frac{1}{2} \left (\frac {q_r} {q_z} \right)^2
    \frac{4\ell (\ell - 1)}{(q_r r)^{2}} e^{-i 2 \varphi}
  \right\}
  (q_r r)^{\ell} e^{i \ell \varphi}
\\ \label{Eq:B_z_antipar}
    \tilde{B}_z(\mathbf r)
&=&
    -i \frac {B_ 0} {2^{\ell-1} (\ell-1)!}
    \frac {q_r} {q_z}
    (q_r r)^{\ell-1} e^{i (\ell-1) \varphi}
\,.
\end{eqnarray}
\end{subequations}

For the magnetic field, the dependence on $r$ is strongly modified
by the combination of polarization and topological charge. In the
parallel class it behaves similar to the electric field; in
particular the in-plane components dominate and vary as $(q_r
r)^\ell$. In the antiparallel class for $\ell=1$, like in the case
of the electric field, the $z$-component becomes dominant behaving
as $(q_r r)^{(\ell-1)}$. Even more interesting, for antiparallel
beams with $\ell \ge 2$ there are second-order terms in the ratio
$(q_r/q_z)$ proportional to $(q_r r)^{\ell-2}$ in the in-plane
components. Being solutions of the full wave equation, for Bessel
beams $(q_r/q_z)$ may take any value, and such terms may become
important under strong focussing. This is the reason why one cannot
write the interaction Hamiltonian only in terms of electric fields,
as can be done in the TL gauge for beams in the parallel class
\cite{quinteiro2015formulation}.


\section{Formulating the interaction in terms of electric
and magnetic fields} \label{Sec:Formulation}

We derive the TL-matter Hamiltonian in the Poincar\'e gauge. The
starting point are the general formulas \cite{cohen1997pho}
\begin{subequations}
\begin{eqnarray}
\label{Eq:Pot-A-Poinc}
    \mathbf A(\mathbf r,t)
&=&
    -\int_0^1 \, du \, u \,\mathbf r \times \mathbf B(u\mathbf r,t)
\\
\label{Eq:Pot-U-Poinc}
    U(\mathbf r,t)
&=&
    -\int_0^1 \, du \,\mathbf r \cdot \mathbf E(u\mathbf r,t)
\,.
\end{eqnarray}
\end{subequations}
These potentials are inserted in the standard minimal coupling Hamiltonian
\begin{equation}
H =\frac{1}{2m}[\mathbf p - q \mathbf A(\mathbf r,t)]^2 + V(\mathbf r) + qU(\mathbf r,t),
\end{equation}
where $V(\mathbf r)$ denotes a static potential for the particles with charge
$q$ and mass $m$. Assuming that the term proportional to $\mathbf{A}^2$ is
negligible, which is well justified for typical magnetic field strengths in a
light beam, the coupling to the electric field is then described by the
Hamiltonian
\begin{equation}
\label{Eq:Hamiltonian_E}
    H_e = q  U(\mathbf r, t)
\end{equation}
while for the coupling to the magnetic field we obtain
\begin{eqnarray}
\label{Eq:Hamiltonian_B}
    H_m &=& - \frac{q}{2m} \left[ \mathbf p \cdot \mathbf A(\mathbf r,t) +
    \mathbf A(\mathbf r,t) \cdot \mathbf p \right] \,.
\end{eqnarray}
We are interested in the interaction of the beam with flat, nano-sized
structures with radial extensions much smaller than the beam waist located
around $z=0$ and $r=0$. Therefore, we can use the approximate field profiles
of Eqs.~\eqref{Eq:E_Bessel}-\eqref{Eq:B_Bessel_antipar} and take the
propagating phase factor at $z=0$. We adopt the convention $\mathbf r_\perp =
r \hat{\mathbf r}$.

\subsection{Electric interaction}

We first consider the interaction with the electric field, for which we
already derived the TL gauge for the parallel class. Using the electric field
from Sec.~\ref{Sec:OV} we can easily evaluate the integral in
Eq.~\eqref{Eq:Pot-U-Poinc}. Note, that the transverse components of the
parallel and antiparallel beams have the same $r$-dependence $(q_r
r)^{|\ell|}$, while for the $z$ component $(q_r r)^{|\ell+\sigma|}$. In
total, according to Eq.~\eqref{Eq:Hamiltonian_E} the electric Hamiltonian for
the interaction with a particle with charge $q$ is
\begin{eqnarray} \label{Eq:U_final}
    H_e
&=&
    - \frac{1}{|\ell|+1}  \, q \mathbf r_{\perp} \cdot \mathbf
    E_\perp(\mathbf r_\perp, t)
\nonumber \\
&&
    - \frac{1}{|\ell+\sigma|+1} \, q z \, E_z(\mathbf r_\perp, t)
\,,
\end{eqnarray}
where $\mathbf E_\perp = (E_x, E_y)$, and we call $q \mathbf r_{\perp}$ the
in-plane dipole moment $\mathbf d$, although the interaction is actually
multipolar. We also want to stress the appearance of the prefactors due to
the vortex structure of the field. Equation (\ref{Eq:U_final}) is in
agreement with our previous results \cite{quinteiro2015formulation}, but is
valid for the parallel and antiparallel classes.

Next, we want to check, whether the approximation $z=0$ assuming a flat
structure holds. For this, we include the next order of the Taylor expansion
in $z$ of the field, $\vE(\vr,t)\approx\frac{1}{2} \tilde{\vE}(\mathbf
r_\perp)[1+iq_z z] e^{-i\omega t} + \textrm{c.c.}$. Also for this field
Eq.~\eqref{Eq:Pot-U-Poinc} can be readily evaluated giving rise to a second
order contribution to the interaction Hamiltonian according to
\begin{eqnarray} \label{Eq:U_final_2}
    H_e^{(2)}
&\sim&
    - q_z \frac{1}{|\ell|+1}  \,  q z \mathbf r_{\perp} \cdot \mathbf
    E_\perp(\mathbf r_\perp, t)
\nonumber \\
&&
    - q_z \frac{q}{|\ell+\sigma|+2}  \, q z^2 \, E_z(\mathbf r_\perp, t)
\,,
\end{eqnarray}
As example we could think of a planar nanostructure excited by optical
fields. For instance a disk-shaped QD \cite{quinteiro2009ele1} with $2-5$~nm
height impinged at normal incidence by a light pulse of $q_z = 2\pi/\lambda=
0.01$nm$^{-1}$ yields $q_z |z|<0.05$. Indeed, we see that the first term
dominates, which is ensured by the condition $q_z |z| \ll 1$. It is worth
mentioning that there are situations in which higher orders are required; for
instance, due to the parity of the initial $\Psi_i$ and final $\Psi_f$ states
involved in the optical transition, the term $\bra{\Psi_f} \{- [1/(|\ell|+1)]
\, \mathbf r_\perp \cdot \mathbf E_\perp(\mathbf r_\perp, t)\} \ket{\Psi_i}$
might be zero.

\subsection{Magnetic (orbital) interaction}

Now we turn to the interaction induced by the magnetic parts of the field.

Using Eqs.~\eqref{Eq:Pot-A-Poinc} and \eqref{Eq:Hamiltonian_B} as well as the
identities $\mathbf p \cdot (\mathbf r \times \mathbf B) = (\mathbf p \times
\mathbf r) \cdot \mathbf B)$ and $(\mathbf r \times \mathbf B) \cdot \mathbf
p = \mathbf B \cdot ( \mathbf p \times \mathbf r)$, we obtain
\begin{eqnarray}
\label{Eq:H_m}
    H_m
    &=&    \frac{q}{2m} (\mathbf p \times \mathbf r) \cdot \int_0^1 \, du \, u
    \,\mathbf B(u \mathbf r_\perp,t) \nonumber \\
    &&    + \frac{q}{2m}     \int_0^1 \, du \, u \,\mathbf B(u\mathbf r_\perp,t)\,
    \cdot \left(\mathbf p \times \mathbf r  \right) \,.
\end{eqnarray}
Using that the commutator $[\mathbf B(\mathbf r,t), (\mathbf p \times \mathbf
r)]$ is small (see Appendix \ref{App:MI}), we put together both terms in
Eq.~(\ref{Eq:H_m}) simplifying our interaction to
\begin{eqnarray}
\label{Eq:H_m1}
    H_m &=&  \frac{q}{m}   \left[  \int_0^1 \, du \, u \,\mathbf B(u\mathbf r_\perp,t)\,
    \right] \cdot \left(\mathbf p \times \mathbf r  \right) \,.
\end{eqnarray}
Inserting the magnetic fields from Sec.~\ref{Sec:OV}, the evaluation of the
integral is straightforward, resulting in
\begin{eqnarray}
\label{Eq:H_m_v2}
    H_m &=&     \frac{2}{|\ell|+2-j}     \mathbf B_\perp(\mathbf r_\perp,t) \cdot
    \left[\frac{q}{2m} (\mathbf p \times \mathbf r)\right] \nonumber \\
&&    + \frac{2}{|\ell+\sigma|+2}     B_z(\mathbf r_\perp,t) \hat{\mathbf z} \cdot
\left[\frac{q}{2m} (\mathbf p \times \mathbf r)\right]
\,,
\end{eqnarray}
with $j=2$ for antiparallel beams with $\ell \ge 2$ and $j=0$ otherwise.

We will call $[-(q/2m)(\mathbf p \times \mathbf r)]$ the magnetic moment
$\mathbf m_B$ keeping in mind that the interaction terms is a multipolar
interaction. Of course, in the simplest case of homogeneous fields $\mathbf
B(0,t)$ one recovers the well-known magnetic-dipole interaction $H_m=-\mathbf
m_B \cdot \mathbf B(0,t)$.


\section{Analyzing the Hamiltonian}
\label{Sec:diss}

For compactness and to reinforce the resemblance with well-known formulas
used for smooth fields, we may define effective fields
\begin{eqnarray*}
\label{eq:H_compact}
    \mathbf E^{\mathrm {eff}}(\mathbf r_\perp, t) &=&  \frac{1}{|\ell|+1} \, \mathbf E_\perp(\mathbf r_\perp, t) \nonumber \\
    &&   + \frac{1}{|\ell+\sigma|+1} \, E_z(\mathbf r_\perp, t) \hat{\mathbf z} \nonumber \\
    \mathbf B^{\mathrm {eff}}(\mathbf r_\perp, t) &=&  \frac{2}{|\ell|+2-j} \mathbf B_\perp(\mathbf r_\perp,t) \nonumber \\
    &&     + \frac{2}{|\ell+\sigma|+2}  B_z(\mathbf r_\perp,t) \hat{\mathbf z} \,,
\end{eqnarray*}
that allow us to write the complete Hamiltonian in an appealing form
\begin{eqnarray} \label{Eq:H_final}
    H &=& \frac{\mathbf p^2}{2m}  + V(\mathbf r) - \mathbf E^{\mathrm {eff}}(\mathbf r_\perp, t) \cdot \mathbf d \nonumber \\
        && - \mathbf B^{\mathrm {eff}}(\mathbf r_\perp, t) \cdot \mathbf m_B \,,
\end{eqnarray}
for it is {\it local} depending solely in the position vector
$\mathbf r$ and is {\it intuitive} reminding the well-known
dipole-moment interactions. We next discuss some features of our
gauge as well some separate cases to illustrate the effects of
TL-matter interaction.

\subsection{Comparison to the multipolar expansion:}

Using simple arguments the Hamiltonian Eq.\ (\ref{Eq:H_final}) can be
compared to the multipolar expansion \cite{cohen1997pho}. When expanding the
electric field terms in Eq.~(\ref{eq:H_compact}) we regain the lowest order
multipolar terms. For example if $\ell = 1$ the transverse electric field
$E_{\perp}(\mathbf r,t) \propto (q_r r)$ and the interaction $H_e =
-\frac{1}{2} \mathbf r_{\perp} \cdot \mathbf E_{\perp}(\mathbf r,t) \propto
r^2$ is electric quadrupolar in $r$. If the OAM is increased to $\ell = 2$,
the interaction becomes $H_e \propto r^3$, an electric octupole. This is in
agreement with our previous findings \cite{quinteiro2015formulation}.


\subsection{Transverse ($xy$)-components:}
In the paraxial approximation and also in most cases of interest, the
transverse components of the fields play the major role. In this case, the
Hamiltonian reduces to the simple form
\begin{eqnarray} \label{Eq:H_transverse}
    H^{\perp}_{\textrm{int}} &=&  - \frac{1}{|\ell|+1} \, \mathbf E_\perp(\mathbf r_\perp, t) \cdot \mathbf d  \nonumber \\
                    &&   - \frac{2}{|\ell|+2-j} \mathbf B_\perp(\mathbf r_\perp,t) \cdot \mathbf
                    m_B \, .
\end{eqnarray}
For the parallel class it can be shown that the electric component always
dominates. For this, we remind the reader that the magnetic-$2^n$-pole
interaction is weaker than the electric-$2^n$-pole interaction. This is
clearly the case for homogeneous fields: with the use of $\langle\mathbf
p\rangle = -i(m/\hbar) \langle[\mathbf r,H_0]\rangle$ the magnetic-dipole
interaction $-\mathbf m_B \cdot \mathbf B(0,t) \propto |({\mathbf r} \times
{\mathbf p})| \propto \langle r \rangle ^2$ while the electric-dipole
interaction $-\mathbf d \cdot \mathbf E(0,t)\propto \langle r \rangle$. For
parallel beams of TL, the $r$-dependence of magnetic and electric fields is
the same $[(q_r r)^\ell]$, and the argument for homogeneous fields can be
used to assert that the strongest interaction is the electric one.

For the antiparallel class, one has to be more careful and reconsider the
fields given in Eqs.~\eqref{Eq:E_Bessel} and \eqref{Eq:B_Bessel_antipar}. For
$\ell$=1 both fields are proportional to $r$ and for the same arguments as
above, the electric field dominates. More interesting is the case $\ell=2$.
On the one hand, the electric interaction is $\mathbf d \cdot \mathbf
E(\mathbf r,t)\propto r (q_r\, r)^2 \propto r^3$, an electric octupole. On
the other hand, the magnetic field is constant (no singularity), and its
interaction is thus magnetic dipolar with $\mathbf B(\mathbf r,t) \cdot
(\mathbf p \times \mathbf r) \propto \langle r \rangle ^2$. This indicates
that the magnetic interaction dominates close to $r=0$. Our conclusion is
supported by Zurita's {\it et al.} \cite{zurita2002multipolar} study on the
interaction of spherical QD with focused azimuthally polarized beam, where
the transition rate is larger for the magnetic interaction. For the cases
$\ell>2$, we also find that the magnetic field, which is proportional to
$r^{\ell-2}$ overcomes the electric field, proportional to $r^{\ell}$.

In contrast to the TL gauge, the Hamiltonian in Eq.~\eqref{Eq:H_transverse}
accounts for both electric and magnetic field and is therefore able to
capture all cases of handedness of polarization $\sigma$ and topological
charge $\ell$.

\subsection{Longitudinal ($z$)-component:}
The interaction Hamiltonian for the $z$-components of the field can be
written as follows
\begin{eqnarray} \label{Eq:H_perp}
    H^{z}_{\mathrm{int}} &=&  - \frac{1}{|\ell+\sigma|+1} \, E_z(\mathbf r_\perp, t)  d_z  \nonumber \\
                    &&   - \frac{2}{|\ell+\sigma|+2} B_z(\mathbf r_\perp,t) m_{B,z}\,.
\end{eqnarray}
For the $z$-components of the field, using the same arguments as above, the
electric field dominates over the magnetic one in all cases.

%
\section{Extension to radially and azimuthally polarized fields}
\label{Sec:Ext}

Radially and azimuthally polarized beams can be built as a superposition of
two antiparallel twisted light beams having $\{\ell=1,\sigma=-1\}$ and
$\{\ell=-1,\sigma=1\}$. Azimuthally polarized fields are given by the sum of
these beams. In the approximation of small $r$ the fields read
\begin{eqnarray}
\label{eq:E_az}
   E^{\mathrm{(az)}}_\varphi
&=&
    E_0  (q_r r) e^{i(q_z z - \omega t)} +\textrm{c.c.} \nonumber \\
   E^{\mathrm{(az)}}_r
&=&
   E^{\mathrm{(az)}}_z = 0 \,,
\end{eqnarray}
and
\begin{eqnarray}
\label{eq:B_az}
     B^{\mathrm{(az)}}_r (\vr,t) &=&   -  B_0 (q_r r) e^{i(q_z z - \omega t)} +\textrm{c.c.}\nonumber \\
     B^{\mathrm{(az)}}_\varphi  (\vr,t) &=&     0 \nonumber \\
     B^{\mathrm{(az)}}_z  (\vr,t) &=&   - 2  i B_0 \frac{q_r}{q_z} e^{i(q_z z - \omega t)} +\textrm{c.c.} \,.
\end{eqnarray}
Likewise the radially polarized fields given by the difference of the two
antiparallel beams are
\begin{eqnarray}
\label{eq:E_rad}
     E^{\mathrm{(rad)}}_r  (\vr,t) &=&   i E_0  (q_r r) e^{i(q_z z - \omega t)} +\textrm{c.c.} \nonumber \\
     E^{\mathrm{(rad)}}_\varphi  (\vr,t) &=&     0 \nonumber \\
     E^{\mathrm{(rad)}}_z  (\vr,t)  &=&     - 2 \frac{q_r}{q_z} E_0  e^{i(q_z z - \omega t)} +\textrm{c.c.} \,,
\end{eqnarray}
and
\begin{eqnarray}
\label{eq:B_rad}
     B^{\mathrm{(rad)}}_\varphi  (\vr,t)  &=&     i B_0 \left[ 1 +  \left(\frac{q_r}{q_z}\right)^2 \right]
     (q_r r) e^{i(q_z z - \omega t)} +\textrm{c.c.} \nonumber \\
     B^{\mathrm{(rad)}}_r  (\vr,t)  &=&     B^{\mathrm{(rad)}}_z = 0 \,.
\end{eqnarray}
Evidently, for both types of fields all in-plane components, varying as $(q_r
r)$, vanish at the origin. In contrast, at $r=0$ the azimuthally polarized
beam is characterized by a non-vanishing $z$-component of the magnetic field
while the radially polarized beam exhibits a non-vanishing $z$-component of
the electric field. Thus, close to the beam center both fields are dominated
by their longitudinal contributions.

Due to the particular mixture of polarization and topological charge the
prefactors containing $|\ell+\sigma|$ and $|\ell|$ in Eqs.~(\ref{Eq:U_final})
and (\ref{Eq:H_m_v2}) are the same for each single $\{\ell,\sigma\}$ field.
Thus, radially and azimuthally polarized beams can be directly used in the
Hamiltonian expression Eq.~(\ref{Eq:H_final}). Working out the interaction
terms for the $z$-components of the fields, we find for the radially
polarized field
\begin{eqnarray}
\label{Eq:H_Ra}
    H_e
&=&
    - E^{\mathrm{(rad)}}_z(\mathbf r_\perp, t)  \,  d_z
\,,
\end{eqnarray}
and for the azimuthally polarized field
\begin{eqnarray}
\label{Eq:H_Az}
    H_m
&=&
    - B^{\mathrm{(az)}}_z(\mathbf r_\perp,t) \, m_{B,z}
\,.
\end{eqnarray}
Note that there is no prefactor, since all $z$-components have no phase singularity.

\section{Conclusions}
\label{Sec:Conc}

We have revisited the mathematical formulation of the TL-matter interaction
close to the phase singularity. In follow-up to the TL gauge
\cite{quinteiro2015formulation}, we have extended the gauge-invariant
formulation by applying a transformation to the Poincar\'e gauge to both
classes (parallel and antiparallel) of TL as well as to azimuthally and
radially polarized beams. The Hamiltonian includes both electric and magnetic
interaction, which is important, because for a particular combination of
orbital and spin momenta the TL-matter interaction is dominated by the
magnetic field, a very uncommon situation in optics. An important advantage
of the Hamiltonian is that it is written {\it solely} in terms of fields,
overcoming issues of gauge invariances. The expression is both local and
intuitive, resembling well-known formulas used to study smooth light fields.

\section{Acknowledgment}

G.~F.~Quinteiro would like to thank the Argentine research agency Agencia
Nacional de Promocion Cientifica y Tecnologica and the Institut f\"ur
Festk\"orpertheorie of WWU M\"unster (Germany) for financial support.

\appendix

\section{Potential and fields for Bessel beams}
\label{App:Bessel}

Using again the separation of the propagating phase from the mode function
according to $\vA(\vr,t)=\frac{1}{2} \tilde{\vA}(\vr)e^{i(q_z z-\omega t)} +
\textrm{c.c.}$, the vector potential of Bessel beams is
\cite{quinteiro2015formulation}
\begin{eqnarray}
    \tilde{\vA}(\mathbf{r}) &=& A_0 \,\left[ \mathbf{e}_{\sigma} J_{\ell}(q_r r)
    e^{i \ell \varphi}  \right. \nonumber \\
    &&     \left.     - i \,\sigma \mathbf{e}_z
    \frac{q_r}{q_z} J_{\ell + \sigma}(q_r r)     e^{i (\ell + \sigma) \varphi} \right] \,,
\label{eq:A}
\end{eqnarray}
with frequency $\omega$, wave vectors $q_z$ and $q_r$, related by
$q_z^2+q_r^2 = (n\omega/c)^2$, $A_0$ the amplitude and $n$ being the index of
refraction of the medium. $J_{\ell}$ denotes the Bessel function and
$\mathbf{e}_{\sigma}=(\mathbf{e}_x + i \sigma \mathbf{e}_y)$ is the
polarization with $\mathbf{e}_{x}$ ($\mathbf{e}_{y}$) is the unit vector in
$x$ ($y$) direction. The scalar potential can be chosen to $\Phi(\vr,t)=0$,
such that the fields are calculated in the standard way via $\mathbf E = -
\frac{\partial}{\partial t} \mathbf A$ and $\mathbf B = \nabla \times A$,
which for the electric field yields
\begin{subequations}
\label{Eq:E_Bessel_full}
\begin{eqnarray}
\label{Eq:EBx_full}
    \tilde{E}_x(\mathbf{r})
&=&
    i E_0  J_{\ell}(q_r r) e^{i \ell \varphi} \,,
\\
\label{Eq:EBy_full}
    \tilde{E}_y(\mathbf{r})
&=&
    -\sigma  E_0 J_{\ell}(q_r r) e^{i \ell \varphi} \,,
\\
\label{Eq:EBz_full}
    \tilde{E}_z(\mathbf{r})
&=&
    \sigma E_0 \frac{q_r}{q_z} J_{\ell + \sigma}(q_r r) e^{i (\ell + \sigma) \varphi} \,,
\end{eqnarray}
\end{subequations}
with $E_0 = \omega A_0$. The magnetic field reads
\begin{subequations}
\label{Eq:B_Bessel_full}
\begin{eqnarray}
\label{Eq:B_x_full}
    \tilde{B}_x(\mathbf{r})
&=&
    \sigma B_0 \biggl[ \Bigl( 1 + \frac{q_r^2}{2q_z^2} - \frac{q_r^2}{2q_z^2}
    e^{i 2 \sigma\varphi} \Bigr) J_\ell(q_r r) e^{i \ell \varphi} \nonumber \\
    && + \frac{q_r^2}{2q_z^2} (\ell + \sigma) \frac{2}{q_r r}
    J_{\ell + \sigma}(q_r r) e^{i (\ell + 2 \sigma) \varphi} \biggr] \,,
\\ \label{Eq:B_y_full}
    \tilde{B}_y(\mathbf{r})
&=&
    i B_0 \biggl[ \Bigl( 1 + \frac{q_r^2}{2q_z^2} + \frac{q_r^2}{2q_z^2}
    e^{i 2 \sigma\varphi} \Bigr) J_\ell(q_r r) e^{i \ell \varphi} \nonumber \\
    && - \frac{q_r^2}{2q_z^2} (\ell + \sigma) \frac{2}{q_r r}
    J_{\ell + \sigma}(q_r r) e^{i (\ell + 2 \sigma) \varphi} \biggr]    \,,
\\ \label{Eq:B_z_full}
    \tilde{B}_z(\mathbf{r})
&=&
    -i B_0 \frac{q_r}{q_z} J_{\ell + \sigma}(q_r r) e^{i (\ell + \sigma) \varphi}
\,,
\end{eqnarray}
\end{subequations}
with $B_0 = q_z A_0 = (q_z / \omega) E_0$. The behavior close to the beam
center given in Sec.~\ref{Sec:OV} is obtained from the expansion
\begin{equation}
J_\nu(z) = \frac{z^\nu}{2^\nu \nu !} \left[ 1 - \frac{z^2}{4(\nu +1)} + \dots \right]
\end{equation}
valid for $\nu \ge 0$ and the relation $J_{-\nu}(z) = (-1)^\nu J_{\nu}(x)$.

\section{Commutator $[\mathbf B(\mathbf r,t), (\mathbf p \times \mathbf r)]$}
\label{App:MI}

Considering linear materials and
that $\mathbf p \times \mathbf r = - \mathbf r \times \mathbf p$
\begin{eqnarray}
\label{Eq:comm1}
    [\mathbf B(\mathbf r,t), (\mathbf p \times \mathbf r)]
&=&
 \mathbf r \cdot [\mathbf p \times \mathbf B(\mathbf r,t)]
\nonumber \\
&=&
    -i \hbar \mathbf r \cdot [\nabla \times \mathbf B(\mathbf r,t)]
\nonumber \\
&=&
    -i \hbar \mu \epsilon  \mathbf r \cdot
    \frac{\partial \mathbf E(\mathbf r,t)}{\partial t}
\,,
\end{eqnarray}
where in the last line we used Ampere-Maxwell's equation and the
fact that the current $\mathbf j(\mathbf r,t)$ --source of $E$ and
$B$-- is far away and can be disregarded. For a monochromatic field
$\partial_t \mathbf E(\mathbf r,t) = - i \omega \mathbf E(\mathbf
r,t)$ and $\mu \epsilon = 1/c^2$, then the correction to the
magnetic Hamiltonian $H_m$ is
\begin{eqnarray}
\label{Eq:comm2}
    \Delta H_m &=& \int_0^1 du\, u\, \frac{q}{2m} [\mathbf B(u\mathbf r,t),
    (\mathbf p \times \mathbf r)] \nonumber \\
&=&
    -\frac{q}{2}
    \frac{\hbar\omega}{m c^2}
    \int_0^1 du\, u^2 \mathbf r \cdot \mathbf E(u\mathbf r,t)
\,.
\end{eqnarray}
Thus, the correction has a structure similar to the electric Hamiltonion,
however with a prefactor $\hbar\omega/(m c^2) \simeq 10^{-5}$. It can
therefore be safely disregarded.


\bibliography{std,twistedlight_v1}

\begin{thebibliography}{34}%
\makeatletter
\providecommand \@ifxundefined [1]{%
 \@ifx{#1\undefined}
}%
\providecommand \@ifnum [1]{%
 \ifnum #1\expandafter \@firstoftwo
 \else \expandafter \@secondoftwo
 \fi
}%
\providecommand \@ifx [1]{%
 \ifx #1\expandafter \@firstoftwo
 \else \expandafter \@secondoftwo
 \fi
}%
\providecommand \natexlab [1]{#1}%
\providecommand \enquote  [1]{``#1''}%
\providecommand \bibnamefont  [1]{#1}%
\providecommand \bibfnamefont [1]{#1}%
\providecommand \citenamefont [1]{#1}%
\providecommand \href@noop [0]{\@secondoftwo}%
\providecommand \href [0]{\begingroup \@sanitize@url \@href}%
\providecommand \@href[1]{\@@startlink{#1}\@@href}%
\providecommand \@@href[1]{\endgroup#1\@@endlink}%
\providecommand \@sanitize@url [0]{\catcode `\\12\catcode `\$12\catcode
  `\&12\catcode `\#12\catcode `\^12\catcode `\_12\catcode `\%12\relax}%
\providecommand \@@startlink[1]{}%
\providecommand \@@endlink[0]{}%
\providecommand \url  [0]{\begingroup\@sanitize@url \@url }%
\providecommand \@url [1]{\endgroup\@href {#1}{\urlprefix }}%
\providecommand \urlprefix  [0]{URL }%
\providecommand \Eprint [0]{\href }%
\providecommand \doibase [0]{http://dx.doi.org/}%
\providecommand \selectlanguage [0]{\@gobble}%
\providecommand \bibinfo  [0]{\@secondoftwo}%
\providecommand \bibfield  [0]{\@secondoftwo}%
\providecommand \translation [1]{[#1]}%
\providecommand \BibitemOpen [0]{}%
\providecommand \bibitemStop [0]{}%
\providecommand \bibitemNoStop [0]{.\EOS\space}%
\providecommand \EOS [0]{\spacefactor3000\relax}%
\providecommand \BibitemShut  [1]{\csname bibitem#1\endcsname}%
\let\auto@bib@innerbib\@empty
\bibitem [{\citenamefont {Andrews}(2008)}]{andrews2011str}%
  \BibitemOpen
  \bibfield  {author} {\bibinfo {author} {\bibfnamefont {D.~L.}\ \bibnamefont
  {Andrews}},\ }\href@noop {} {\emph {\bibinfo {title} {Structured light and
  its applications: An introduction to phase-structured beams and nanoscale
  optical forces}}}\ (\bibinfo  {publisher} {Academic Press},\ \bibinfo {year}
  {2008})\BibitemShut {NoStop}%
\bibitem [{\citenamefont {Ballantine}\ \emph {et~al.}(2016)\citenamefont
  {Ballantine}, \citenamefont {Donegan},\ and\ \citenamefont
  {Eastham}}]{ballantine2016there}%
  \BibitemOpen
  \bibfield  {author} {\bibinfo {author} {\bibfnamefont {K.~E.}\ \bibnamefont
  {Ballantine}}, \bibinfo {author} {\bibfnamefont {J.~F.}\ \bibnamefont
  {Donegan}}, \ and\ \bibinfo {author} {\bibfnamefont {P.~R.}\ \bibnamefont
  {Eastham}},\ }\href@noop {} {\bibfield  {journal} {\bibinfo  {journal}
  {Science Advances}\ }\textbf {\bibinfo {volume} {2}},\ \bibinfo {pages}
  {e1501748} (\bibinfo {year} {2016})}\BibitemShut {NoStop}%
\bibitem [{\citenamefont {Yamane}\ \emph {et~al.}(2012)\citenamefont {Yamane},
  \citenamefont {Toda},\ and\ \citenamefont {Morita}}]{yamane2012ultrashort}%
  \BibitemOpen
  \bibfield  {author} {\bibinfo {author} {\bibfnamefont {K.}~\bibnamefont
  {Yamane}}, \bibinfo {author} {\bibfnamefont {Y.}~\bibnamefont {Toda}}, \ and\
  \bibinfo {author} {\bibfnamefont {R.}~\bibnamefont {Morita}},\ }\href@noop {}
  {\bibfield  {journal} {\bibinfo  {journal} {Optics express}\ }\textbf
  {\bibinfo {volume} {20}},\ \bibinfo {pages} {18986} (\bibinfo {year}
  {2012})}\BibitemShut {NoStop}%
\bibitem [{\citenamefont {K{\"o}ksal}\ and\ \citenamefont
  {Berakdar}(2012)}]{koksal2012cha}%
  \BibitemOpen
  \bibfield  {author} {\bibinfo {author} {\bibfnamefont {K.}~\bibnamefont
  {K{\"o}ksal}}\ and\ \bibinfo {author} {\bibfnamefont {J.}~\bibnamefont
  {Berakdar}},\ }\href@noop {} {\bibfield  {journal} {\bibinfo  {journal}
  {Phys.\ Rev.\ A}\ }\textbf {\bibinfo {volume} {86}},\ \bibinfo {pages}
  {063812} (\bibinfo {year} {2012})}\BibitemShut {NoStop}%
\bibitem [{\citenamefont {Surzhykov}\ \emph {et~al.}(2015)\citenamefont
  {Surzhykov}, \citenamefont {Seipt}, \citenamefont {Serbo},\ and\
  \citenamefont {Fritzsche}}]{surzhykov2015interaction}%
  \BibitemOpen
  \bibfield  {author} {\bibinfo {author} {\bibfnamefont {A.}~\bibnamefont
  {Surzhykov}}, \bibinfo {author} {\bibfnamefont {D.}~\bibnamefont {Seipt}},
  \bibinfo {author} {\bibfnamefont {V.}~\bibnamefont {Serbo}}, \ and\ \bibinfo
  {author} {\bibfnamefont {S.}~\bibnamefont {Fritzsche}},\ }\href@noop {}
  {\bibfield  {journal} {\bibinfo  {journal} {Phys.\ Rev.\ A}\ }\textbf
  {\bibinfo {volume} {91}},\ \bibinfo {pages} {013403} (\bibinfo {year}
  {2015})}\BibitemShut {NoStop}%
\bibitem [{\citenamefont {W{\"a}tzel}\ \emph {et~al.}(2016)\citenamefont
  {W{\"a}tzel}, \citenamefont {Pavlyukh}, \citenamefont {Sch{\"a}ffer},\ and\
  \citenamefont {Berakdar}}]{watzel2016optical}%
  \BibitemOpen
  \bibfield  {author} {\bibinfo {author} {\bibfnamefont {J.}~\bibnamefont
  {W{\"a}tzel}}, \bibinfo {author} {\bibfnamefont {Y.}~\bibnamefont
  {Pavlyukh}}, \bibinfo {author} {\bibfnamefont {A.}~\bibnamefont
  {Sch{\"a}ffer}}, \ and\ \bibinfo {author} {\bibfnamefont {J.}~\bibnamefont
  {Berakdar}},\ }\href@noop {} {\bibfield  {journal} {\bibinfo  {journal}
  {Carbon}\ }\textbf {\bibinfo {volume} {99}},\ \bibinfo {pages} {439}
  (\bibinfo {year} {2016})}\BibitemShut {NoStop}%
\bibitem [{\citenamefont {Schmiegelow}\ and\ \citenamefont
  {Schmidt-Kaler}(2012)}]{schmiegelow2012light}%
  \BibitemOpen
  \bibfield  {author} {\bibinfo {author} {\bibfnamefont {C.~T.}\ \bibnamefont
  {Schmiegelow}}\ and\ \bibinfo {author} {\bibfnamefont {F.}~\bibnamefont
  {Schmidt-Kaler}},\ }\href@noop {} {\bibfield  {journal} {\bibinfo  {journal}
  {The European Physical Journal D}\ }\textbf {\bibinfo {volume} {66}},\
  \bibinfo {pages} {1} (\bibinfo {year} {2012})}\BibitemShut {NoStop}%
\bibitem [{\citenamefont {Peshkov}\ \emph {et~al.}(2015)\citenamefont
  {Peshkov}, \citenamefont {Fritzsche},\ and\ \citenamefont
  {Surzhykov}}]{peshkov2015ionization}%
  \BibitemOpen
  \bibfield  {author} {\bibinfo {author} {\bibfnamefont {A.}~\bibnamefont
  {Peshkov}}, \bibinfo {author} {\bibfnamefont {S.}~\bibnamefont {Fritzsche}},
  \ and\ \bibinfo {author} {\bibfnamefont {A.}~\bibnamefont {Surzhykov}},\
  }\href@noop {} {\bibfield  {journal} {\bibinfo  {journal} {Physical Review
  A}\ }\textbf {\bibinfo {volume} {92}},\ \bibinfo {pages} {043415} (\bibinfo
  {year} {2015})}\BibitemShut {NoStop}%
\bibitem [{\citenamefont {Bhowmik}\ \emph {et~al.}(2016)\citenamefont
  {Bhowmik}, \citenamefont {Mondal}, \citenamefont {Majumder},\ and\
  \citenamefont {Deb}}]{bhowmik2016interaction}%
  \BibitemOpen
  \bibfield  {author} {\bibinfo {author} {\bibfnamefont {A.}~\bibnamefont
  {Bhowmik}}, \bibinfo {author} {\bibfnamefont {P.~K.}\ \bibnamefont {Mondal}},
  \bibinfo {author} {\bibfnamefont {S.}~\bibnamefont {Majumder}}, \ and\
  \bibinfo {author} {\bibfnamefont {B.}~\bibnamefont {Deb}},\ }\href@noop {}
  {\bibfield  {journal} {\bibinfo  {journal} {Physical Review A}\ }\textbf
  {\bibinfo {volume} {93}},\ \bibinfo {pages} {063852} (\bibinfo {year}
  {2016})}\BibitemShut {NoStop}%
\bibitem [{\citenamefont {Quinteiro}\ and\ \citenamefont
  {Tamborenea}(2009{\natexlab{a}})}]{quinteiro2009the}%
  \BibitemOpen
  \bibfield  {author} {\bibinfo {author} {\bibfnamefont {G.~F.}\ \bibnamefont
  {Quinteiro}}\ and\ \bibinfo {author} {\bibfnamefont {P.~I.}\ \bibnamefont
  {Tamborenea}},\ }\href@noop {} {\bibfield  {journal} {\bibinfo  {journal}
  {Europhys.\ Lett.}\ }\textbf {\bibinfo {volume} {85}},\ \bibinfo {pages}
  {47001} (\bibinfo {year} {2009}{\natexlab{a}})}\BibitemShut {NoStop}%
\bibitem [{\citenamefont {Shigematsu}\ \emph {et~al.}(2013)\citenamefont
  {Shigematsu}, \citenamefont {Toda}, \citenamefont {Yamane},\ and\
  \citenamefont {Morita}}]{shigematsu2013orb}%
  \BibitemOpen
  \bibfield  {author} {\bibinfo {author} {\bibfnamefont {K.}~\bibnamefont
  {Shigematsu}}, \bibinfo {author} {\bibfnamefont {Y.}~\bibnamefont {Toda}},
  \bibinfo {author} {\bibfnamefont {K.}~\bibnamefont {Yamane}}, \ and\ \bibinfo
  {author} {\bibfnamefont {R.}~\bibnamefont {Morita}},\ }\href@noop {}
  {\bibfield  {journal} {\bibinfo  {journal} {Jpn.\ J.\ Appl.\ Phys.}\ }\textbf
  {\bibinfo {volume} {52}},\ \bibinfo {pages} {08JL08} (\bibinfo {year}
  {2013})}\BibitemShut {NoStop}%
\bibitem [{\citenamefont {Clayburn}\ \emph {et~al.}(2013)\citenamefont
  {Clayburn}, \citenamefont {McCarter}, \citenamefont {Dreiling}, \citenamefont
  {Poelker}, \citenamefont {Ryan},\ and\ \citenamefont
  {Gay}}]{clayburn2013sea}%
  \BibitemOpen
  \bibfield  {author} {\bibinfo {author} {\bibfnamefont {N.~B.}\ \bibnamefont
  {Clayburn}}, \bibinfo {author} {\bibfnamefont {J.~L.}\ \bibnamefont
  {McCarter}}, \bibinfo {author} {\bibfnamefont {J.~M.}\ \bibnamefont
  {Dreiling}}, \bibinfo {author} {\bibfnamefont {M.}~\bibnamefont {Poelker}},
  \bibinfo {author} {\bibfnamefont {D.~M.}\ \bibnamefont {Ryan}}, \ and\
  \bibinfo {author} {\bibfnamefont {T.~J.}\ \bibnamefont {Gay}},\ }\href@noop
  {} {\bibfield  {journal} {\bibinfo  {journal} {Phys.\ Rev.\ {\rm B}}\
  }\textbf {\bibinfo {volume} {87}},\ \bibinfo {pages} {035204} (\bibinfo
  {year} {2013})}\BibitemShut {NoStop}%
\bibitem [{\citenamefont {Noyan}\ and\ \citenamefont
  {Kikkawa}(2015)}]{noyan2015time}%
  \BibitemOpen
  \bibfield  {author} {\bibinfo {author} {\bibfnamefont {M.~A.}\ \bibnamefont
  {Noyan}}\ and\ \bibinfo {author} {\bibfnamefont {J.~M.}\ \bibnamefont
  {Kikkawa}},\ }\href@noop {} {\bibfield  {journal} {\bibinfo  {journal}
  {Appl.\ Phys.\ Lett.}\ }\textbf {\bibinfo {volume} {107}},\ \bibinfo {pages}
  {032406} (\bibinfo {year} {2015})}\BibitemShut {NoStop}%
\bibitem [{\citenamefont {Shintani}\ \emph {et~al.}(2016)\citenamefont
  {Shintani}, \citenamefont {Taguchi}, \citenamefont {Tanaka},\ and\
  \citenamefont {Kawaguchi}}]{shintani2016spin}%
  \BibitemOpen
  \bibfield  {author} {\bibinfo {author} {\bibfnamefont {K.}~\bibnamefont
  {Shintani}}, \bibinfo {author} {\bibfnamefont {K.}~\bibnamefont {Taguchi}},
  \bibinfo {author} {\bibfnamefont {Y.}~\bibnamefont {Tanaka}}, \ and\ \bibinfo
  {author} {\bibfnamefont {Y.}~\bibnamefont {Kawaguchi}},\ }\href@noop {}
  {\bibfield  {journal} {\bibinfo  {journal} {Phys.\ Rev.\ {\rm B}}\ }\textbf
  {\bibinfo {volume} {93}},\ \bibinfo {pages} {195415} (\bibinfo {year}
  {2016})}\BibitemShut {NoStop}%
\bibitem [{\citenamefont {Ko{\c{c}}}\ and\ \citenamefont
  {K{\"o}ksal}(2015)}]{kocc2015quantum}%
  \BibitemOpen
  \bibfield  {author} {\bibinfo {author} {\bibfnamefont {F.}~\bibnamefont
  {Ko{\c{c}}}}\ and\ \bibinfo {author} {\bibfnamefont {K.}~\bibnamefont
  {K{\"o}ksal}},\ }\href@noop {} {\bibfield  {journal} {\bibinfo  {journal}
  {Superlattices and Microstructures}\ }\textbf {\bibinfo {volume} {85}},\
  \bibinfo {pages} {599} (\bibinfo {year} {2015})}\BibitemShut {NoStop}%
\bibitem [{\citenamefont {Omatsu}\ \emph {et~al.}(2010)\citenamefont {Omatsu},
  \citenamefont {Chujo}, \citenamefont {Miyamoto}, \citenamefont {Okida},
  \citenamefont {Nakamura}, \citenamefont {Aoki},\ and\ \citenamefont
  {Morita}}]{omatsu2010metal}%
  \BibitemOpen
  \bibfield  {author} {\bibinfo {author} {\bibfnamefont {T.}~\bibnamefont
  {Omatsu}}, \bibinfo {author} {\bibfnamefont {K.}~\bibnamefont {Chujo}},
  \bibinfo {author} {\bibfnamefont {K.}~\bibnamefont {Miyamoto}}, \bibinfo
  {author} {\bibfnamefont {M.}~\bibnamefont {Okida}}, \bibinfo {author}
  {\bibfnamefont {K.}~\bibnamefont {Nakamura}}, \bibinfo {author}
  {\bibfnamefont {N.}~\bibnamefont {Aoki}}, \ and\ \bibinfo {author}
  {\bibfnamefont {R.}~\bibnamefont {Morita}},\ }\href@noop {} {\bibfield
  {journal} {\bibinfo  {journal} {Optics express}\ }\textbf {\bibinfo {volume}
  {18}},\ \bibinfo {pages} {17967} (\bibinfo {year} {2010})}\BibitemShut
  {NoStop}%
\bibitem [{\citenamefont {Spinello}\ \emph {et~al.}(2016)\citenamefont
  {Spinello}, \citenamefont {Someda}, \citenamefont {Ravanelli}, \citenamefont
  {Mari}, \citenamefont {Parisi}, \citenamefont {Tamburini}, \citenamefont
  {Romanato}, \citenamefont {Coassini},\ and\ \citenamefont
  {Oldoni}}]{spinello2016radio}%
  \BibitemOpen
  \bibfield  {author} {\bibinfo {author} {\bibfnamefont {F.}~\bibnamefont
  {Spinello}}, \bibinfo {author} {\bibfnamefont {C.~G.}\ \bibnamefont
  {Someda}}, \bibinfo {author} {\bibfnamefont {R.~A.}\ \bibnamefont
  {Ravanelli}}, \bibinfo {author} {\bibfnamefont {E.}~\bibnamefont {Mari}},
  \bibinfo {author} {\bibfnamefont {G.}~\bibnamefont {Parisi}}, \bibinfo
  {author} {\bibfnamefont {F.}~\bibnamefont {Tamburini}}, \bibinfo {author}
  {\bibfnamefont {F.}~\bibnamefont {Romanato}}, \bibinfo {author}
  {\bibfnamefont {P.}~\bibnamefont {Coassini}}, \ and\ \bibinfo {author}
  {\bibfnamefont {M.}~\bibnamefont {Oldoni}},\ }\href@noop {} {\bibfield
  {journal} {\bibinfo  {journal} {AEU-International Journal of Electronics and
  Communications}\ }\textbf {\bibinfo {volume} {70}},\ \bibinfo {pages} {990}
  (\bibinfo {year} {2016})}\BibitemShut {NoStop}%
\bibitem [{\citenamefont {Zhang}\ \emph {et~al.}(2016)\citenamefont {Zhang},
  \citenamefont {Yu}, \citenamefont {Zhang}, \citenamefont {Xu}, \citenamefont
  {Xu},\ and\ \citenamefont {Wang}}]{zhang2016self}%
  \BibitemOpen
  \bibfield  {author} {\bibinfo {author} {\bibfnamefont {Y.}~\bibnamefont
  {Zhang}}, \bibinfo {author} {\bibfnamefont {H.}~\bibnamefont {Yu}}, \bibinfo
  {author} {\bibfnamefont {H.}~\bibnamefont {Zhang}}, \bibinfo {author}
  {\bibfnamefont {X.}~\bibnamefont {Xu}}, \bibinfo {author} {\bibfnamefont
  {J.}~\bibnamefont {Xu}}, \ and\ \bibinfo {author} {\bibfnamefont
  {J.}~\bibnamefont {Wang}},\ }\href@noop {} {\bibfield  {journal} {\bibinfo
  {journal} {Optics Express}\ }\textbf {\bibinfo {volume} {24}},\ \bibinfo
  {pages} {5514} (\bibinfo {year} {2016})}\BibitemShut {NoStop}%
\bibitem [{\citenamefont {Abulikemu}\ \emph {et~al.}(2016)\citenamefont
  {Abulikemu}, \citenamefont {Yusufu}, \citenamefont {Mamuti}, \citenamefont
  {Araki}, \citenamefont {Miyamoto},\ and\ \citenamefont
  {Omatsu}}]{abulikemu2016octave}%
  \BibitemOpen
  \bibfield  {author} {\bibinfo {author} {\bibfnamefont {A.}~\bibnamefont
  {Abulikemu}}, \bibinfo {author} {\bibfnamefont {T.}~\bibnamefont {Yusufu}},
  \bibinfo {author} {\bibfnamefont {R.}~\bibnamefont {Mamuti}}, \bibinfo
  {author} {\bibfnamefont {S.}~\bibnamefont {Araki}}, \bibinfo {author}
  {\bibfnamefont {K.}~\bibnamefont {Miyamoto}}, \ and\ \bibinfo {author}
  {\bibfnamefont {T.}~\bibnamefont {Omatsu}},\ }\href@noop {} {\bibfield
  {journal} {\bibinfo  {journal} {Optics Express}\ }\textbf {\bibinfo {volume}
  {24}},\ \bibinfo {pages} {15204} (\bibinfo {year} {2016})}\BibitemShut
  {NoStop}%
\bibitem [{\citenamefont {Miao}\ \emph {et~al.}(2016)\citenamefont {Miao},
  \citenamefont {Zhang}, \citenamefont {Sun}, \citenamefont {Walasik},
  \citenamefont {Longhi}, \citenamefont {Litchinitser},\ and\ \citenamefont
  {Feng}}]{miao2016orbital}%
  \BibitemOpen
  \bibfield  {author} {\bibinfo {author} {\bibfnamefont {P.}~\bibnamefont
  {Miao}}, \bibinfo {author} {\bibfnamefont {Z.}~\bibnamefont {Zhang}},
  \bibinfo {author} {\bibfnamefont {J.}~\bibnamefont {Sun}}, \bibinfo {author}
  {\bibfnamefont {W.}~\bibnamefont {Walasik}}, \bibinfo {author} {\bibfnamefont
  {S.}~\bibnamefont {Longhi}}, \bibinfo {author} {\bibfnamefont {N.~M.}\
  \bibnamefont {Litchinitser}}, \ and\ \bibinfo {author} {\bibfnamefont
  {L.}~\bibnamefont {Feng}},\ }\href@noop {} {\bibfield  {journal} {\bibinfo
  {journal} {Science}\ }\textbf {\bibinfo {volume} {353}},\ \bibinfo {pages}
  {464} (\bibinfo {year} {2016})}\BibitemShut {NoStop}%
\bibitem [{\citenamefont {Quinteiro}\ and\ \citenamefont
  {Kuhn}(2014)}]{quinteiro2014light}%
  \BibitemOpen
  \bibfield  {author} {\bibinfo {author} {\bibfnamefont {G.~F.}\ \bibnamefont
  {Quinteiro}}\ and\ \bibinfo {author} {\bibfnamefont {T.}~\bibnamefont
  {Kuhn}},\ }\href@noop {} {\bibfield  {journal} {\bibinfo  {journal} {Phys.\
  Rev.\ {\rm B}}\ }\textbf {\bibinfo {volume} {90}},\ \bibinfo {pages} {115401}
  (\bibinfo {year} {2014})}\BibitemShut {NoStop}%
\bibitem [{\citenamefont {Woerdemann}\ \emph {et~al.}(2013)\citenamefont
  {Woerdemann}, \citenamefont {Alpmann}, \citenamefont {Esseling},\ and\
  \citenamefont {Denz}}]{woerdemann2013advanced}%
  \BibitemOpen
  \bibfield  {author} {\bibinfo {author} {\bibfnamefont {M.}~\bibnamefont
  {Woerdemann}}, \bibinfo {author} {\bibfnamefont {C.}~\bibnamefont {Alpmann}},
  \bibinfo {author} {\bibfnamefont {M.}~\bibnamefont {Esseling}}, \ and\
  \bibinfo {author} {\bibfnamefont {C.}~\bibnamefont {Denz}},\ }\href@noop {}
  {\bibfield  {journal} {\bibinfo  {journal} {Laser \& Photonics Reviews}\
  }\textbf {\bibinfo {volume} {7}},\ \bibinfo {pages} {839} (\bibinfo {year}
  {2013})}\BibitemShut {NoStop}%
\bibitem [{\citenamefont {Saito}\ \emph {et~al.}(2008)\citenamefont {Saito},
  \citenamefont {Kobayashi}, \citenamefont {Hiraga}, \citenamefont {Fujita},
  \citenamefont {Kawano}, \citenamefont {Smith}, \citenamefont {Inouye},\ and\
  \citenamefont {Kawata}}]{saito2008zpolarization}%
  \BibitemOpen
  \bibfield  {author} {\bibinfo {author} {\bibfnamefont {Y.}~\bibnamefont
  {Saito}}, \bibinfo {author} {\bibfnamefont {M.}~\bibnamefont {Kobayashi}},
  \bibinfo {author} {\bibfnamefont {D.}~\bibnamefont {Hiraga}}, \bibinfo
  {author} {\bibfnamefont {K.}~\bibnamefont {Fujita}}, \bibinfo {author}
  {\bibfnamefont {S.}~\bibnamefont {Kawano}}, \bibinfo {author} {\bibfnamefont
  {N.~I.}\ \bibnamefont {Smith}}, \bibinfo {author} {\bibfnamefont
  {Y.}~\bibnamefont {Inouye}}, \ and\ \bibinfo {author} {\bibfnamefont
  {S.}~\bibnamefont {Kawata}},\ }\href@noop {} {\bibfield  {journal} {\bibinfo
  {journal} {J.\ Raman Spectroscopy}\ }\textbf {\bibinfo {volume} {39}},\
  \bibinfo {pages} {1643} (\bibinfo {year} {2008})}\BibitemShut {NoStop}%
\bibitem [{\citenamefont {Wang}\ \emph {et~al.}(2008)\citenamefont {Wang},
  \citenamefont {Shi}, \citenamefont {Lukyanchuk}, \citenamefont {Sheppard},\
  and\ \citenamefont {Chong}}]{wang2008creation}%
  \BibitemOpen
  \bibfield  {author} {\bibinfo {author} {\bibfnamefont {H.}~\bibnamefont
  {Wang}}, \bibinfo {author} {\bibfnamefont {L.}~\bibnamefont {Shi}}, \bibinfo
  {author} {\bibfnamefont {B.}~\bibnamefont {Lukyanchuk}}, \bibinfo {author}
  {\bibfnamefont {C.}~\bibnamefont {Sheppard}}, \ and\ \bibinfo {author}
  {\bibfnamefont {C.~T.}\ \bibnamefont {Chong}},\ }\href@noop {} {\bibfield
  {journal} {\bibinfo  {journal} {Nat. Photon.}\ }\textbf {\bibinfo {volume}
  {2}},\ \bibinfo {pages} {501} (\bibinfo {year} {2008})}\BibitemShut {NoStop}%
\bibitem [{\citenamefont {Meier}\ \emph {et~al.}(2007)\citenamefont {Meier},
  \citenamefont {Romano},\ and\ \citenamefont {Feurer}}]{meier2007material}%
  \BibitemOpen
  \bibfield  {author} {\bibinfo {author} {\bibfnamefont {M.}~\bibnamefont
  {Meier}}, \bibinfo {author} {\bibfnamefont {V.}~\bibnamefont {Romano}}, \
  and\ \bibinfo {author} {\bibfnamefont {T.}~\bibnamefont {Feurer}},\
  }\href@noop {} {\bibfield  {journal} {\bibinfo  {journal} {Appl.\ Phys.\ A}\
  }\textbf {\bibinfo {volume} {86}},\ \bibinfo {pages} {329} (\bibinfo {year}
  {2007})}\BibitemShut {NoStop}%
\bibitem [{\citenamefont {Zhan}(2004)}]{zhan2004trapping}%
  \BibitemOpen
  \bibfield  {author} {\bibinfo {author} {\bibfnamefont {Q.}~\bibnamefont
  {Zhan}},\ }\href@noop {} {\bibfield  {journal} {\bibinfo  {journal} {Optics
  Express}\ }\textbf {\bibinfo {volume} {12}},\ \bibinfo {pages} {3377}
  (\bibinfo {year} {2004})}\BibitemShut {NoStop}%
\bibitem [{\citenamefont {Sbierski}\ \emph {et~al.}(2013)\citenamefont
  {Sbierski}, \citenamefont {Quinteiro},\ and\ \citenamefont
  {Tamborenea}}]{sbierski2013twisted}%
  \BibitemOpen
  \bibfield  {author} {\bibinfo {author} {\bibfnamefont {B.}~\bibnamefont
  {Sbierski}}, \bibinfo {author} {\bibfnamefont {G.}~\bibnamefont {Quinteiro}},
  \ and\ \bibinfo {author} {\bibfnamefont {P.}~\bibnamefont {Tamborenea}},\
  }\href@noop {} {\bibfield  {journal} {\bibinfo  {journal} {J.\ Phys.\ Cond.\
  Matter}\ }\textbf {\bibinfo {volume} {25}},\ \bibinfo {pages} {385301}
  (\bibinfo {year} {2013})}\BibitemShut {NoStop}%
\bibitem [{\citenamefont {Gabriel}\ \emph {et~al.}(2011)\citenamefont
  {Gabriel}, \citenamefont {Aiello}, \citenamefont {Zhong}, \citenamefont
  {Euser}, \citenamefont {Joly}, \citenamefont {Banzer}, \citenamefont
  {F{\"o}rtsch}, \citenamefont {Elser}, \citenamefont {Andersen}, \citenamefont
  {Marquardt} \emph {et~al.}}]{gabriel2011entangling}%
  \BibitemOpen
  \bibfield  {author} {\bibinfo {author} {\bibfnamefont {C.}~\bibnamefont
  {Gabriel}}, \bibinfo {author} {\bibfnamefont {A.}~\bibnamefont {Aiello}},
  \bibinfo {author} {\bibfnamefont {W.}~\bibnamefont {Zhong}}, \bibinfo
  {author} {\bibfnamefont {T.}~\bibnamefont {Euser}}, \bibinfo {author}
  {\bibfnamefont {N.}~\bibnamefont {Joly}}, \bibinfo {author} {\bibfnamefont
  {P.}~\bibnamefont {Banzer}}, \bibinfo {author} {\bibfnamefont
  {M.}~\bibnamefont {F{\"o}rtsch}}, \bibinfo {author} {\bibfnamefont
  {D.}~\bibnamefont {Elser}}, \bibinfo {author} {\bibfnamefont {U.~L.}\
  \bibnamefont {Andersen}}, \bibinfo {author} {\bibfnamefont {C.}~\bibnamefont
  {Marquardt}},  \emph {et~al.},\ }\href@noop {} {\bibfield  {journal}
  {\bibinfo  {journal} {Phys.\ Rev.\ Lett.}\ }\textbf {\bibinfo {volume}
  {106}},\ \bibinfo {pages} {060502} (\bibinfo {year} {2011})}\BibitemShut
  {NoStop}%
\bibitem [{\citenamefont {Dennis}\ \emph {et~al.}(2009)\citenamefont {Dennis},
  \citenamefont {O'Holleran},\ and\ \citenamefont
  {Padgett}}]{dennis2009singular}%
  \BibitemOpen
  \bibfield  {author} {\bibinfo {author} {\bibfnamefont {M.~R.}\ \bibnamefont
  {Dennis}}, \bibinfo {author} {\bibfnamefont {K.}~\bibnamefont {O'Holleran}},
  \ and\ \bibinfo {author} {\bibfnamefont {M.~J.}\ \bibnamefont {Padgett}},\
  }\href@noop {} {\bibfield  {journal} {\bibinfo  {journal} {Progress in
  Optics}\ }\textbf {\bibinfo {volume} {53}},\ \bibinfo {pages} {293} (\bibinfo
  {year} {2009})}\BibitemShut {NoStop}%
\bibitem [{\citenamefont {Quinteiro}\ \emph {et~al.}(2015)\citenamefont
  {Quinteiro}, \citenamefont {Reiter},\ and\ \citenamefont
  {Kuhn}}]{quinteiro2015formulation}%
  \BibitemOpen
  \bibfield  {author} {\bibinfo {author} {\bibfnamefont {G.}~\bibnamefont
  {Quinteiro}}, \bibinfo {author} {\bibfnamefont {D.}~\bibnamefont {Reiter}}, \
  and\ \bibinfo {author} {\bibfnamefont {T.}~\bibnamefont {Kuhn}},\ }\href@noop
  {} {\bibfield  {journal} {\bibinfo  {journal} {Phys.\ Rev.\ A}\ }\textbf
  {\bibinfo {volume} {91}},\ \bibinfo {pages} {033808} (\bibinfo {year}
  {2015})}\BibitemShut {NoStop}%
\bibitem [{\citenamefont {Volke-Sepulveda}\ \emph {et~al.}(2002)\citenamefont
  {Volke-Sepulveda}, \citenamefont {Garc{\'e}s-Ch{\'a}vez}, \citenamefont
  {Ch{\'a}vez-Cerda}, \citenamefont {Arlt},\ and\ \citenamefont
  {Dholakia}}]{volke2002orb}%
  \BibitemOpen
  \bibfield  {author} {\bibinfo {author} {\bibfnamefont {K.}~\bibnamefont
  {Volke-Sepulveda}}, \bibinfo {author} {\bibfnamefont {V.}~\bibnamefont
  {Garc{\'e}s-Ch{\'a}vez}}, \bibinfo {author} {\bibfnamefont {S.}~\bibnamefont
  {Ch{\'a}vez-Cerda}}, \bibinfo {author} {\bibfnamefont {J.}~\bibnamefont
  {Arlt}}, \ and\ \bibinfo {author} {\bibfnamefont {K.}~\bibnamefont
  {Dholakia}},\ }\href@noop {} {\bibfield  {journal} {\bibinfo  {journal} {J.\
  Opt.\ B}\ }\textbf {\bibinfo {volume} {4}},\ \bibinfo {pages} {S82} (\bibinfo
  {year} {2002})}\BibitemShut {NoStop}%
\bibitem [{\citenamefont {Cohen-Tannoudji}\ \emph {et~al.}(1989)\citenamefont
  {Cohen-Tannoudji}, \citenamefont {Dupont-Roc},\ and\ \citenamefont
  {Grynberg}}]{cohen1997pho}%
  \BibitemOpen
  \bibfield  {author} {\bibinfo {author} {\bibfnamefont {C.}~\bibnamefont
  {Cohen-Tannoudji}}, \bibinfo {author} {\bibfnamefont {J.}~\bibnamefont
  {Dupont-Roc}}, \ and\ \bibinfo {author} {\bibfnamefont {G.}~\bibnamefont
  {Grynberg}},\ }\href@noop {} {\emph {\bibinfo {title} {Photons and Atoms:
  Introduction to Quantum Electrodynamics}}}\ (\bibinfo  {publisher} {Wiley},\
  \bibinfo {year} {1989})\BibitemShut {NoStop}%
\bibitem [{\citenamefont {Quinteiro}\ and\ \citenamefont
  {Tamborenea}(2009{\natexlab{b}})}]{quinteiro2009ele1}%
  \BibitemOpen
  \bibfield  {author} {\bibinfo {author} {\bibfnamefont {G.~F.}\ \bibnamefont
  {Quinteiro}}\ and\ \bibinfo {author} {\bibfnamefont {P.~I.}\ \bibnamefont
  {Tamborenea}},\ }\href@noop {} {\bibfield  {journal} {\bibinfo  {journal}
  {Phys.\ Rev.\ {\rm B}}\ }\textbf {\bibinfo {volume} {79}},\ \bibinfo {pages}
  {155450} (\bibinfo {year} {2009}{\natexlab{b}})}\BibitemShut {NoStop}%
\bibitem [{\citenamefont {Zurita-S{\'a}nchez}\ and\ \citenamefont
  {Novotny}(2002)}]{zurita2002multipolar}%
  \BibitemOpen
  \bibfield  {author} {\bibinfo {author} {\bibfnamefont {J.~R.}\ \bibnamefont
  {Zurita-S{\'a}nchez}}\ and\ \bibinfo {author} {\bibfnamefont
  {L.}~\bibnamefont {Novotny}},\ }\href@noop {} {\bibfield  {journal} {\bibinfo
   {journal} {J.\ Opt.\ Soc.\ Am.\ B}\ }\textbf {\bibinfo {volume} {19}},\
  \bibinfo {pages} {2722} (\bibinfo {year} {2002})}\BibitemShut {NoStop}%
\end{thebibliography}%

\end{document}